\documentclass[11pt,twoside]{article}

\usepackage{asp2004}
\usepackage{epsf}
\usepackage{psfig}
\usepackage{lscape}

\def\Msun{\>{\rm M_{\odot}}}

\begin{document}
\title{Spitzer Search for Mid-IR excesses Around Five DAZs}   
\author{John H. Debes}   
\affil{Department of Terrestrial Magnetism, Carnegie Institution of Washington, Washington, D.C. 20015}
\author{Steinn Sigurdsson}
\affil{Department of Astronomy \& Astrophysics, Pennsylvania State
University, University Park, PA 16802}    
\begin{abstract} 
Hydrogen atmosphere white dwarfs with metals, so-called DAZs,
require external accretion of material to explain the presence of weak metal
line absorption in their photospheres.  The source of this material is currently uknown, but could 
come from the interstellar medium, unseen companions, or relic planetesimals
from asteroid belt or Kuiper belt analogues.  Mid-infrared photometry
of these white dwarfs provide additional information to solve the mystery of 
this accretion and to look for evidence of planetary systems that have 
survived post main sequence evolution.  We present {\em Spitzer} IRAC 
photometry of five DAZs and search for excesses due to unseen companions or
 circumstellar dust disks.  Three of our targets show unexpected 
{\em deficits} in flux, which we tentatively attribute to absorption due to
SiO and CO.
\end{abstract}



\section{Introduction}
White dwarfs have long been used to probe the low mass end of the IMF to look
for low mass stars and brown dwarfs \citep{probst82,zuckerman92,farihi05}.  With
the advent of more sensitive ground- and space-based imaging at longer wavelengths, the direct detection of substellar objects and planets with a few times
Jupiter's mass is now possible \citep{ignace01,burleigh02,friedrich05,debes05,debes05a}.

Searching a subset of white dwarfs that harbor markers for 
substellar objects can maximize the return of such a survey.  Nearby hydrogen
white dwarfs with metal line absorption (DAZs) may fit this criterion.  Three
hypotheses have been put forth to explain the presence of DAZs: interstellar
matter (ISM) accretion \citep{dupuis92,dupuis93a,dupuis93b}, 
unseen companion wind accretion \citep{zuckerman03},
 and accretion of 
volatile poor planetesimals \citep{alcock86,debes02,jura03}.  Accretion of 
planetesimals is certainly the explanation for the disks around GD 362 and G~29-38, as well as other candidate dust disks found with near-IR excesses \citep{kilic06}.  It is less clear that is the explanation for those white dwarfs
without any noticeable near-IR excess.

With the launch of {\em Spitzer} an unprecedented sensitivity is now possible
to further constrain the presence of companions in close orbits, as well as
the presence of dusty disks around white dwarfs.  
A large interest in infrared excesses
around white dwarfs in general is evidenced by the many surveys of white
dwarfs with {\em Spitzer} \citep{hansen06,kilic06}.

In this paper we present results of our search with {\em Spitzer} 
of five nearby DAZs with
no known excesses for companions and circumstellar disks. 

\section{Observations}
\begin{table}[t]
\caption{\label{tab:targs} Properties of the Target White Dwarfs}
\begin{center}
\begin{tabular}{lcccccc}
\tableline
\noalign{\smallskip}
 {WD} &  {T$_{eff}$} &  {t$_{cool}$} &  {D} &  {M$_i$} &  {t$_{cool}$+t$_{MS}$} &  {References} \\
&  {(K)} &  {(Gyr)} &  {(pc)} &  {($\Msun$)} &  {(Gyr)} & \\
\noalign{\smallskip}
\tableline
\noalign{\smallskip}
0208+396 & 7310 & 1.4 & 17 & 2.1 & 3.2  & 1  \\
0243-026 & 6820 & 2.3 & 21 & 3.2 & 2.8 & 1 \\
0245+541 & 5280 & 6.9 & 10 & 4.6 & 7.2 & 1 \\
1257+278 & 8540 & 0.9 & 34 & 1.7 & 3.3 & 1 \\
1620-391 & 24406 & 0.1 & 12 & 3.1 & 0.7 & 2,3 \\
\noalign{\smallskip}
\tableline
\noalign{\smallskip}
\end{tabular}
\end{center}

{\small (1) \citet{bergeron01} (2) \citet{bragaglia95} (3) \citet{vanaltena95}}

\end{table}
Table \ref{tab:targs} shows our target DAZs, complete with known T$_{eff}$, log g, distances, and ages.  Cooling ages were taken from the literature and 
initial masses and main sequence lifetimes were calculated by the equations
of \citet{wood92}:

\begin{eqnarray}
M_i & = & 10.4\ln{\frac{M_{WD}}{0.49 \Msun}} \\
t_{MS}& =& 10 M_i^{-2.5}.
\end{eqnarray}
Each target was observed with the four IRAC channels, with nominal
wavelengths of $\sim$3.6, 4.5, 5.6, and 8.0 \micron \citep{fazio04}.  The observations were
carried out in the mapping mode, with 30 random point dithers for each pair of
channels.  At each dither point, the camera integrated for 96 seconds,
for a total of 2880 seconds in each band.  The exception to this was
WD 1620-391, which is a much brighter source.  It had exposure times of 30 s
per dither with 75 dithers for a total integration of 2250 s.

In order to obtain photometry with an accuracy of $\sim$3\%, we followed
the prescription laid out in \citet{reach05b}.  
With the exception of WD 0208+396, we took each target's (BCD) files, divided
them by correction files to negate array position dependent sensitivity,
registered them, and median combined them to obtain a final image using the
MOPEX package \citep{makovoz05}.  We
performed aperture photmetry with a 5 pixel radius ($\sim$6\arcsec), and
used a 10 pixel wide annulus starting 10 pixels away from the source for
background subtraction.  Aperture corrections appropriate for this size
source radius and background annulus were applied, as well as calibration
factors and flux conversions as mentioned in \citet{reach05b}.  Additionally, the 
change in total flux as a function of location within a pixel was accounted
for in the 3.6\micron channel.  No obvious interstellar cirrus was noted
for any of our targets in the 8\micron\ channel.

For two of our targets,  WD 0245+541
and WD 0208+396, we used smaller source apertures of 2 and 3 pixels 
respectively.  For WD 0208+396, there appeared to be a large increase
in flux as a larger radius was used, possibly due to two resolved 
background objects earlier reported in \citet{debes05,debes06}.
For WD 0245+541, several background objects are located at separations 
$>$ 3\arcsec\, so the smallest aperture size was used to minimize their contribution to the flux.  These two cases demostrate why high spatial resolution
sensitive imaging at shorter wavelengths must be performed for a believable 
claim of an excess, in order to rule out the small contribution from redder
background sources.

Due to a large rate of solar protons, the observations of WD 0208+396
 were degraded
by a large flux of cosmic ray events.  In this case median combination may 
mistakenly return a larger flux due to repeated cosmic ray strikes.  For each
channel, each dither point was visually inspected for obvious cosmic rays 
and if one was present near the target the image was rejected. 
 The resulting collection of ``good'' images were median combined to give 
a final image.  In addition to the standard errors in photometry, we added
a 3$\%$ factor to account for the overall uncertainty in the flux calibrations
quoted by \citet{reach05b}.

In order to detect a bona fide excess, one must compare the observed flux with
an expected flux.  In order to compare our observations we took the models
of \citet{bergeron95} as well as the $BVRIJHK$ photometry of \citet{bergeron01}
for four of the five targets.  WD 1620-391 was not part of \citet{bergeron01}'s
survey and so we used a combination of USNOB, Hipparcos, and 2MASS photometry.  We
compared the photometry and predicted models to ensure a good fit, and then 
extrapolated the observed fluxes from the model K flux
out to longer wavelengths under the assumption of blackbody emission using
the model T$_{eff}$.  The error in the extrapolation then comes solely from 
the error in the effective temperature. 
Since most of the observed BLR data fit within 1-$\sigma$ of the 
model, we required that a significant excess (deficit) be
$>$ three times the photometric error above (below) the
calculated model flux in at least one channel.
 
\section{Results}

\subsection{Significant Deficits}
\begin{figure}[t]
\plotone{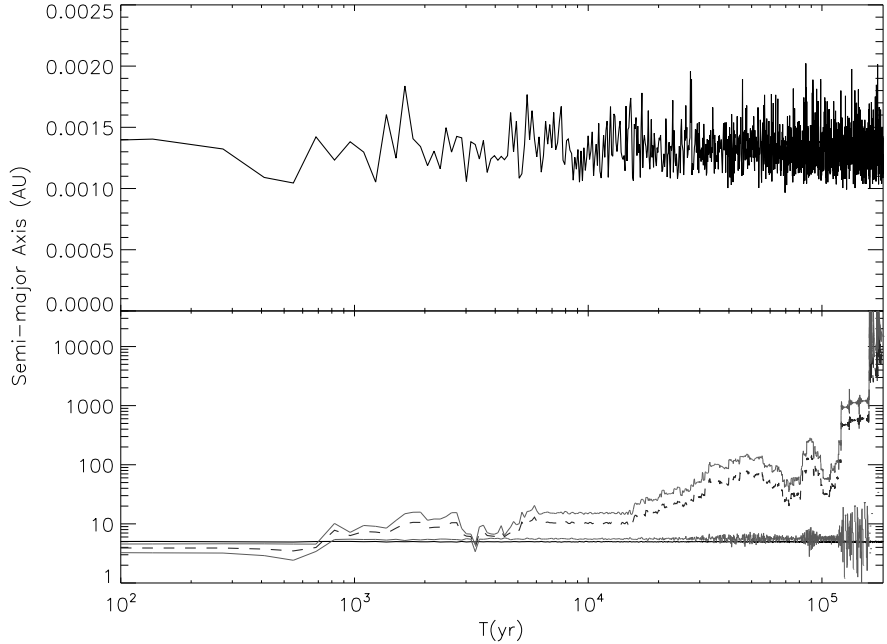}
\caption{\label{fig:wd02} SED of WD 1257+278 compared to predicted
model values and the residuals.  If there is any deficit present, it is
marginal}
\end{figure}

\begin{figure}[t]
\plotone{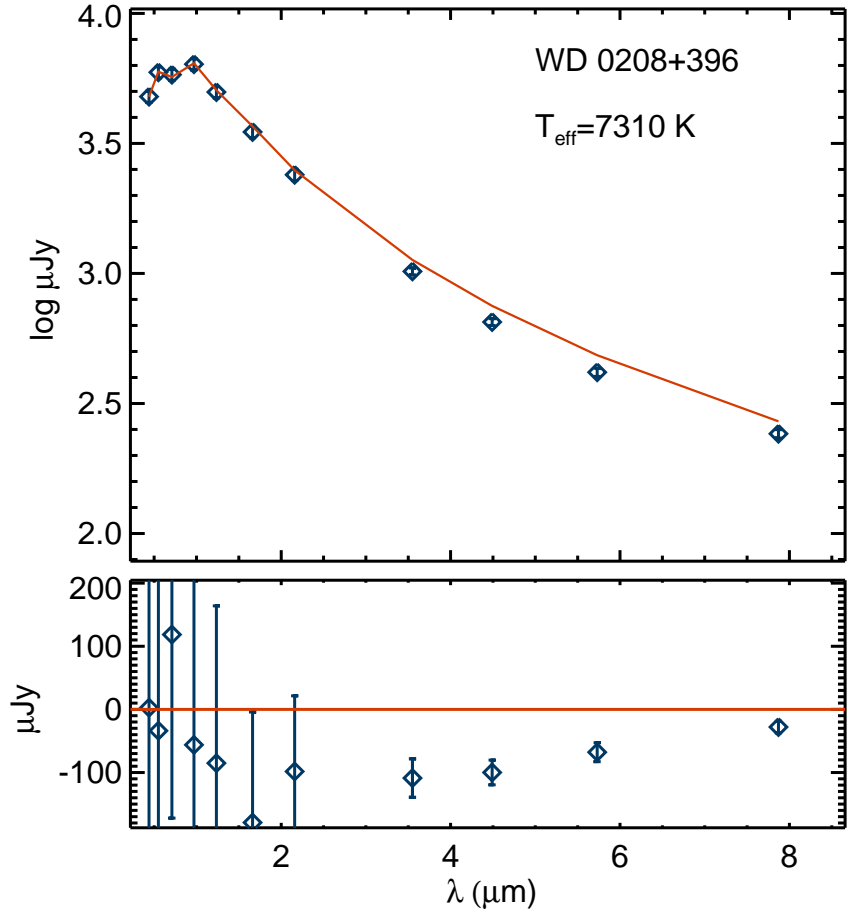}
\caption{\label{fig:wd02a} SED of WD  compared to predicted
model values and the residuals.  There is a deficit between the measured
and expected flux.}
\end{figure}

Figures \ref{fig:wd02a} and \ref{fig:wd02} show the model fluxes and the
measured fluxes for two of our DAZs, as well as the residuals.
Three of the target DAZs show significant deficits in the four channels,
namely WD 0208+396, WD 0245+541, and WD 0243-026.  Their effective temperatures
range from 5280~K to 7310~K, in a similar range to deficits detected in other
DA stars within that temperature range \citep{kilic06}.
We tentatively claim that the source of the opacity is non-LTE absorption by silicon monoxide, with
possibly some contribution from carbon monoxide.
Absorption due to fundamental and overtone rotational-vibrational bands of SiO and CO in late type stars is well known  \cite{cohen}.
The dissociation temperature of SiO is high enough that it could persist at the temperatures of the stars in our sample, and CO could
be present in the cooler stars.  However, the absorption inferred is significantly higher than expected for LTE absorption
at those temperatures and gravities.

We conjecture that SiO is formed above the white dwarf photospheres through photodissociation of silicon dioxide (and any CO present is similarly
formed through photodissociation of carbonates) from refractory dust which sublimates as it is brought down to the the white dwarf surface
through photon drag.
The resulting SiO is formed at low densities above the photosphere, and is far from local thermodynamic equilibrium, with much larger absorption
strengths than inferred from photospheric LTE.

\subsection{Limits to Companions}
For IRAC, very cool substellar objects can be detected as excesses, especially
due to a ``bump'' of flux for brown dwarfs and planets at $\sim$4.5\micron.
While theoretical models predict the 4.5\micron\ flux to be large, observations
of cool brown dwarfs suggest that the spectral models overestimate this flux
by a factor of $\sim$2 \citep{golimowski04}.

In order to place upper limits on the types of unresolved
companions present around our DAZs, we compared predicted IRAC fluxes for
cool brown dwarfs by convolving the IRAC filters with the models of \citet{bsl03} appropriate for the particular age of each target DAZ and its distance.  
For the 4.5\micron\ channel we assumed that the resultant flux was a
factor of two smaller than predicted.  We then compared our 4.5\micron\ 
3$\sigma$ limits to those models in order to determine a mass limit.  

In all cases we improve the 
unresolved companion limits to these objects over \citet{debes05} by a factor
of 2-4.  For WD 0208+395, WD 0243, and WD 1620-391 we rule out all but 
planetary mass objects $<$ 15 M$_{J}$ for separations $<$2.4\arcsec.  
The other two DAZs have limits of $\sim$25 M$_{J}$.  In light of the
fact that some of our sample have anomalous deficits, it should be noted
that these limits may change as the source of the deficit is better understood.

\section{Conclusions}
Five DAZ white dwarfs have been observed with the {\em Spitzer} Telescope for
excesses due to companions or dusty disks.  We find no evidence of excesses, 
but we do find deficits in the flux on the order of 10-20\% between 3.6 and 8\micron.  These deficits require more study, either through spectroscopy of the
white dwarfs or by better modeling of the white dwarf fluxes in the mid-IR.  
We plan to more carefully determine the expected flux of the white dwarfs by
using models that extend into the mid-IR.  The non-LTE behavior of dust grains
impacting a white dwarf atmosphere is a new avenue of study that should provide
further observational diagnostics of the type of material DAZs with
no excess are accreting.

\acknowledgements 
This work is based on observations made with the Spitzer Space Telescope, which is operated by the Jet Propulsion Laboratory, California Institute of Technology under a contract with NASA. Support for this work was provided by NASA through an award issued by JPL/Caltech.


\end{document}